\documentclass[twocolumn,preprintnumbers,amsmath,amssymb]{revtex4-2}
\usepackage[utf8]{inputenc}
\usepackage{enumitem}
\usepackage{epsfig}
\usepackage{graphicx}
\usepackage{dcolumn}
\usepackage{bm}
\usepackage{amsmath, amsfonts}
\usepackage{amssymb}
\usepackage{amsthm}
\usepackage{physics}
\usepackage{comment}
\usepackage{color} 
\usepackage{xcolor}
\usepackage[colorlinks,linkcolor=blue,anchorcolor=blue,urlcolor=blue, citecolor=blue]{hyperref}
\usepackage{siunitx,booktabs}
\usepackage{epstopdf}
\begin{document}
	\title{Advances in device-independent quantum key distribution}
	\author{Víctor Zapatero$^{1,2,3}$}
	\email{vzapatero@com.uvigo.es}
	\author{Tim van Leent$^{4,5}$}
	\author{Rotem Arnon-Friedman$^{6}$}
	\author{Wen-Zhao Liu$^{7,8,9}$}
	\author{Qiang Zhang$^{7,8,9}$}
	\author{Harald Weinfurter$^{4,5,10}$}
	\author{Marcos Curty$^{1,2,3}$}
	\affiliation{$^1$Vigo Quantum Communication Center, University of Vigo, Vigo E-36315, Spain}
	\affiliation{$^2$Escuela de Ingeniería de Telecomunicación, Department of Signal Theory and Communications, University of Vigo, Vigo E-36310, Spain}
	\affiliation{$^3$AtlanTTic Research Center, University of Vigo, Vigo E-36310, Spain}
	\affiliation{$^4$Fakultät für Physik, Ludwig-Maximilians-Universität München, Schellingstr. 4, 80799 München, Germany}
	\affiliation{$^5$Munich Center for Quantum Science and Technology (MCQST), Schellingstr. 4, 80799 München, Germany}
	\affiliation{$^6$Department of Physics of Complex Systems, Weizmann Institute of Science, Rechovot, Israel}
	\affiliation{$^7$Hefei National Research Center for Physical Sciences at the Microscale and School of Physical Sciences, University of Science and Technology of China, Hefei 230026, China}
	\affiliation{$^8$Shanghai Research Center for Quantum Science and CAS Center for Excellence in Quantum Information and Quantum Physics, University of Science and Technology of China, Shanghai 201315, China}
	\affiliation{$^9$Hefei National Laboratory, University of Science and Technology of China, Hefei 230088, China}
	\affiliation{$^{10}$Max-Planck Institut für Quantenoptik, Hans-Kopfermann-Str. 1, 85748 Garching, Germany}
	
\begin{abstract}
	Device-independent quantum key distribution (DI-QKD) provides the gold standard for secure key exchange. Not only it allows for information-theoretic security based on quantum mechanics, but it relaxes the need to physically model the devices, hence fundamentally ruling out many quantum hacking threats to which non-DI QKD systems are vulnerable. In practice though, DI-QKD is very challenging. It relies on the loophole-free violation of a Bell inequality, a task that requires high quality entanglement to be distributed between distant parties and close to perfect quantum measurements, which is hardly achievable with current technology. Notwithstanding, recent theoretical and experimental efforts have led to the first proof-of-principle DI-QKD implementations. In this article, we review the state-of-the-art of DI-QKD by highlighting its main theoretical and experimental achievements, discussing the recent proof-of-principle demonstrations, and emphasizing the existing challenges in the field.
\end{abstract}
\maketitle
\section{Introduction}\label{Introduction}
Quantum key distribution (QKD)~\cite{Pirandola,Xu,Lo} is the remote delivery of secret keys through an insecure channel by using quantum-mechanical information carriers. When combined with the one-time pad encryption scheme~\cite{OTP,Vernam}, QKD allows for information-theoretically secure communications, unbreakable even for an adversary with unlimited computational power. This is in sharp contrast to public-key cryptosystems, threatened by the advent of quantum computers and by the progress of classical computers as well.

Since its conception in 1984~\cite{BB84}, QKD has evolved from a mere theoretical curiosity to a prolific industry at the forefront of quantum technologies. Nowadays, both metropolitan~\cite{Tokyo,Geneve,Cambridge,Yang} and satellite-based~\cite{Liao,Chen} QKD networks are being built, record-breaking transmission distances are being reached over optical fibre~\cite{Boaron,Chen2,Wang}, and QKD services are being supplied by companies around the globe~\cite{IdQuantique,QuantumCTek,Toshiba}. However, despite this success, various important challenges must still be addressed for the widespread application of QKD, related to its security, its performance and its integration with the existing optical communication infrastructure.

In particular, a major difficulty is guaranteeing that the QKD devices behave according to the mathematical models presumed in the security proofs, a problem often termed ``implementation security". Any disparity between these models and the actual operation of the QKD equipment might invalidate the security claims and potentially opens a security loophole. The importance of this problem is evidenced by the amount of quantum hacking attacks reported in the last two decades~\cite{Jain}, which reveal that the single-photon detectors are the Achilles' heel of QKD systems. Remarkably, a breakthrough in this respect was the invention of measurement-device-independent (MDI) QKD~\cite{MDI}, together with its most recent variant, called twin-field (TF) QKD~\cite{TF}. Both solutions remove all security loopholes from the measurement unit but require that the functioning of the QKD transmitter is perfectly characterized. Nevertheless, given the complexity of its multiple optical and electronic components~\cite{Pereira,Pereira2}, exhaustively characterizing a QKD transmitter is still an open task.

In this context, device-independent (DI) QKD~\cite{Mayers,Barrett,Acin,Acin2,Acin3,Masanes,Vazirani,Miller} can be considered the ultimate solution to the problem of implementation security, because it does not require to characterize the internal functioning of any device. The DI feature shows up in the historical Ekert 91 (E91) protocol~\cite{E91}, where a central untrusted source distributes entangled photon pairs between two parties, say Alice and Bob, each provided with a measurement unit. By performing adequate local measurements on the incident photons, the parties can certify the presence of monogamous correlations between their measurement outcomes~\cite{Coffman,Terhal} on the basis of their input-and-output statistics alone. Precisely, the statistics must violate a Bell inequality~\cite{Bell_EPR,nonlocality}, which proves that the outcomescannot be correlated to some local knowledge possibly attributed to an adversary, usually called Eve. 
%

Much progress has been made to quantitatively link the amount of Bell violation to Eve's uncertainty on the parties' measurement outcomes, and to formally prove that a Bell violation enables the extraction of a secure key (see e.g.~\cite{Acin,Acin2,Acin3,Masanes,Vazirani,Miller,Pironio,Pironio_random,Acin4,Tomamichel_uncertainty,Arnon2018,Arnon2019} among other works).

The experimental realization of a so-called ``loophole-free Bell test" is, however, a major challenge. Entanglement has to be distributed among remote observers, which must be capable of performing random measurements at a high speed and with very high efficiency. In a technological tour de force, several loophole-free Bell tests have been performed in the last years~\cite{Hensen,Giustina,Shalm,Rosenfeld,Li}. Notably though, for all studied protocols, the possibility to perform DI-QKD is more stringent than the Bell test itself, because it demands that the Bell violation lie well above what is required for the test.
Notwithstanding, the first proof-of-principle DI-QKD demonstrations have recently been reported~\cite{Nadlinger,Liu,Zhang}.

In this article, we summarize the progress in the field of DI-QKD. First, we revisit the DI paradigm in Section~\ref{Assumptions}, after which we review the main challenges and theoretical progress in Sections~\ref{Protocols} and~\ref{Security}. Thereafter, we turn our attention to the emerging topic of experimental DI-QKD in Section~\ref{Implementations}, with an emphasis on the recent proof-of-principle demonstrations. To finish with, prospects in the field are outlined in Section~\ref{Outlook}.

\section{Security assumptions of the DI setting}\label{Assumptions}
Let us first summarize the assumptions on which DI-QKD relies. (1) Quantum mechanics (QM) is correct. By this, we mean that, in particular, given (only) the quantum-mechanical description of a system and a measurement performed on it, the theory predicts the measurement statistics correctly. Of course, this hypothesis is justified by the perfect agreement between QM and experiments. (2) The parties pre-share a short secret key for the authentication of classical messages. (3) The parties can faithfully generate local randomness. Remarkably, inasmuch as this assumption allows to assure that the measurement settings are ``free" in a precise sense~\cite{Bell2}, it implies the completeness of QM for the prediction of measurement outcomes~\cite{Colbeck} ---a frequently invoked assumption by itself--- via the exclusion of more predictive theories compatible with it. (4) The parties hold faithful classical post-processing units. (5) No unwanted information leakage occurs through the boundaries of the parties' locations.

The above assumptions are a subset of the assumptions made in the non-DI setting. As long as they hold, no further device characterization enters the security analysis of DI-QKD, which provides a significant security upgrade.

 

Arguably though, assumption (5) --- which is also needed in conventional cryptosystems--- might be hard to assure in practice, especially given that entanglement needs to be distributed. In fact, (5) may be replaced by more tenable assumptions modelling the information leakage, to quantify its effect on the secret key rate at the price of slightly undermining the DI feature, as it is done in~\cite{Nadlinger}. Indeed, this approach has also been studied in the non-DI setting~\cite{Lucamarini}.

In addition, (5) restricts Eve’s capability to sabotage the DI-QKD equipment. This is so because, by hypothesis, it prevents a corrupted device from covertly leaking private information, which could in principle be done in very contrived ways~\cite{memory,Xu,malicious}. In this regard, a practical solution to weaken assumption (5) ---and also (4)--- is introduced in~\cite{malicious,malicious2}, combining the use of redundant QKD devices and secret sharing to deal with truly malicious equipment.

Furthermore, another paradigm is being explored to perform DI-QKD with communicating devices ---thus violating assumption (5)--- as long as (i) they are restricted by a post-quantum computational assumption and (ii) Eve does not have direct access to the channel~\cite{Metger}.

\section{DI-QKD protocols and challenges}\label{Protocols}

\begin{figure}[!htbp]
	\centering 
	\includegraphics[width=8cm,height=2.9cm]{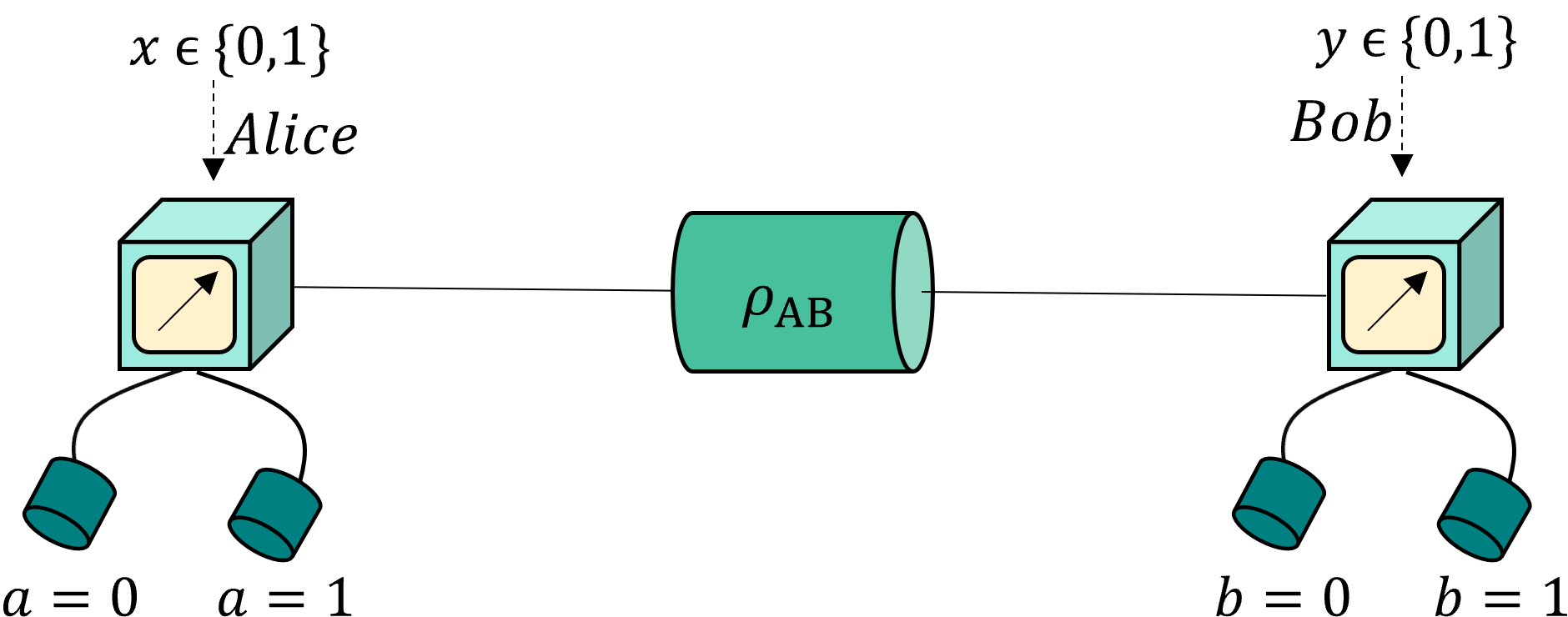}\\
	\caption{CHSH test setup. A central source distributes quantum states $\rho_{\mathrm{AB}}$ to Alice and Bob, who interact with their measurement devices by providing binary inputs ($x$ and $y$) and recording binary outputs ($a$ and $b$) to estimate their CHSH winning probability, $\omega$. As an example, the Tsirelson's bound $\omega=(2+\sqrt{2})/4$ is reached if $\rho_{\mathrm{AB}}=\ketbra{\Phi^{+}}{\Phi^{+}}_{\mathrm{AB}}$ (with $\ket{\Phi^{+}}_{\mathrm{AB}}=(\ket{00}_{\mathrm{AB}}+\ket{11}_{\mathrm{AB}})/\sqrt{2}$) and the inputs $x$ and $y$ determine the measurement of the observables $A_{x}$ and $B_{y}$, where $A_{0}=\sigma_{\rm Z}$, $A_{1}=\sigma_{\rm X}$, $B_{0}=(\sigma_{\rm Z}+\sigma_{\rm X})/\sqrt{2}$ and $B_{1}=(\sigma_{\rm Z}-\sigma_{\rm X})/\sqrt{2}$ ($\sigma_{\rm Z}$ and $\sigma_{\rm X}$ being Pauli matrices). In this ideal example, Alice records output $a$ upon observation of the eigenvalue $(-1)^{a}$, and similarly for Bob.}
	\label{fig:CHSH}
\end{figure}

The central Bell inequality in the bipartite scenario is the Clauser-Horne-Shimony-Holt (CHSH) inequality~\cite{CHSH}, which completely characterizes non-local correlations in the binary inputs and outputs setting~\cite{nonlocality}. A ``CHSH test" can be formulated as a two-party non-local game, as illustrated in Figure~\ref{fig:CHSH}.

In every round of the game, Alice and Bob independently provide random binary inputs $x$ and $y$ to their devices, and win the game if their binary outputs $a$ and $b$ fulfil the CHSH winning condition,
\begin{equation}
a\oplus{}b=x\cdot{}y,
\end{equation}
where ``$\oplus$" and ``$\cdot$" denote addition and multiplication modulo 2, respectively. In this context, the CHSH inequality is an upper bound on the winning probability $\omega$ attainable by any probability distribution $p(a,b|x,y)$ that admits a local description:
\begin{equation}
\omega\leq{}75\%.
\end{equation}
Famously, QM enables a maximum winning probability of $(2+\sqrt{2})/4\approx{}85.4\%$ known as the Tsirelson's bound~\cite{Tsirelson}, attainable by the measurement statistics of Bell pairs upon careful selection of the measurement settings.

Naturally then, a CHSH-based DI-QKD protocol runs sequentially, randomly alternating between key rounds ---where the parties measure their shares in fixed correlated bases--- and test rounds ---where they play the CHSH game to quantify Eve's information on the outcomes of the key rounds---. The key basis is chosen to be one of Alice's test bases, and Bob's device incorporates an extra setting to operate in the same basis. As usual, irrelevant basis choice pairings are dismissed a posteriori.

After the quantum communication phase, the generated raw key material undergoes several standard classical post-processing steps (not necessarily in the following order): sifting ---where the data from unsuitable basis choice pairings is discarded, thus obtaining the sifted keys---, parameter estimation ---where the average score of the CHSH test is calculated---, error correction (EC) ---where, say, Bob computes an estimate of Alice's sifted key with the aid of some public discussion, to ensure that both strings match with a high probability--- and privacy amplification (PA) ---where these latter keys are shrunk into shorter secure bit strings, whose length is prescribed by the parameter estimation step---.

In what follows, we focus on the requirements of implementing CHSH-based DI-QKD. As stated in Section~\ref{Introduction}, any DI-QKD protocol fundamentally relies on the violation of a Bell inequality, and the conclusiveness of the latter is subject to the closure of two main loopholes~\cite{nonlocality}: a locality loophole~\cite{Bell}, and a detection loophole~\cite{Pearle,Clauser,Gisin}. The closure of the locality loophole demands that~\cite{Pironio,nonlocality} (i) the measurement setting choice of one party may not influence the measurement outcome of another party and (ii) the local measurement settings be free choices~\cite{Bell2,Colbeck}. If either condition is not met, a Bell violation admits a local description. By hypothesis though, the no-leakage assumption (5) already implies (i)~\cite{Pironio}, although an orthodox approach to enforce it consists of ensuring a space-like separation between the delivery of one party's input and the return of the other party's output. Similarly, (ii) is guaranteed by assumption (3), which we recall is inevitable in all of cryptography.

On the other hand, the detection loophole demands more attention. In practical Bell tests, signals are lost due to absorption in the channel or device inefficiencies. If lost signals are simply dismissed, an adversary capable of correlating losses to the measurement settings of the parties can easily fake the Bell violation. Consequently, all the signals must be accounted for in the Bell test ---e.g. assigning a specific outcome to undetected signals~\cite{nonlocality}--- which strongly undermines the loss tolerance of DI-QKD as we discuss below.

Hereafter, we refer to the original CHSH-based protocol presented in~\cite{Acin3}. For illustration purposes, let us consider an implementation of the protocol using entangled photon pairs, where the parties close the detection loophole by deterministically mapping lost signals to a fixed photo-detector, as mentioned above. In addition, for benchmarking purposes, we contemplate a typical limited-efficiency model~\cite{Pironio} where each photon is independently lost with a fixed probability, due to channel loss or to the finite detection efficiency of the photo-detectors, $\eta_{\rm det}$. Even if one assumes perfect preparation of Bell states, a positive asymptotic key rate (that is, considering that the parties execute infinitely many protocol rounds) demands that the efficiency of the photo-detectors satisfy $\eta_{\rm det}>92.4\%$ if only detection (but no channel) losses occur~\cite{Pironio}. This can be reduced to $90.9\%$ if Bob undoes the assignment of undetected signals for EC purposes~\cite{Ma}. Complementarily, if only channel (but no detection) losses occur, a positive key rate demands that the user-to-user distance satisfy $L<3.5$ km, considering a typical optical fibre with an attenuation coefficient of 0.2 dB/km at telecom wavelength (see e.g.~\cite{Zapatero}).

While a noticeable progress exists towards close-to-perfect single-photon detection, the constraint on the tolerated channel loss is prohibitive. In fact, if one wants to cover larger distances, the closure of the detection loophole demands the usage of a heralding mechanism, as explained next.\\

\noindent \textbf{Heralding mechanisms} \\

A heralding mechanism is an instrument that informs the parties about the arrival of a photon or the successful distribution of entanglement between them. In this way, one can decouple channel loss from the measurement settings by simply postponing the choice of the latter until the heralding occurs. Naturally, this allows to discard the signals lost in the channel without opening the detection loophole.

\begin{figure}[!htbp]
	\centering 
	\includegraphics[width=8.6cm,height=3.6cm]{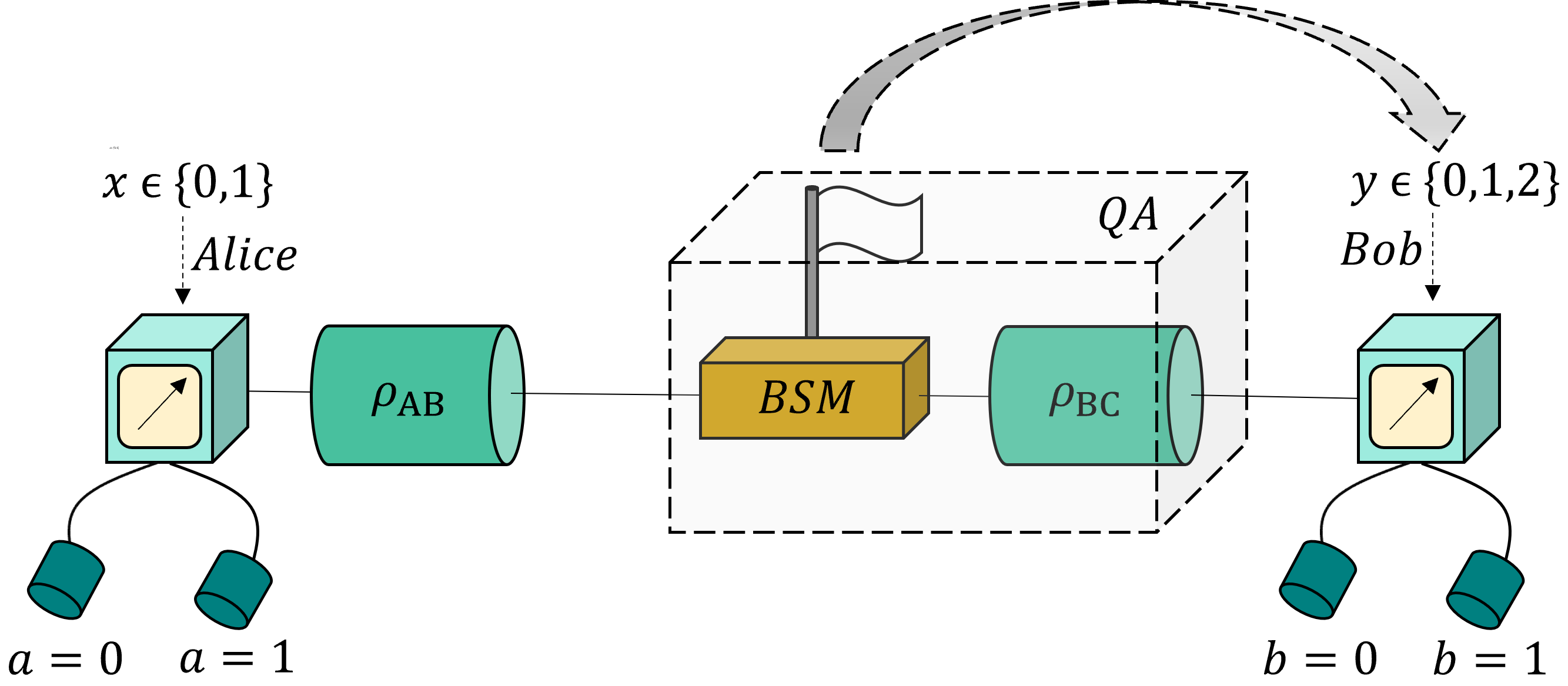} 
	\caption{QA-assisted DI-QKD. One possibility consists} of locating the entanglement source $\rho_{\rm AB}$ at Alice's site, such that only Bob experiences channel loss, and thus he is the one equipped with a QA. Alternatively, one can place $\rho_{\rm AB}$ in the channel and equip both Alice and Bob with QAs. For instance, the QA may be constructed with an auxiliary entanglement source $\rho_{\rm BC}$ and a Bell state measurement (BSM). In the BSM, a travelling photon from $\rho_{\rm AB}$ interferes with a photon from $\rho_{\rm BC}$. Upon success of the BSM, Bob is warned of the arrival of a photon ---which is symbolized by a flag within the QA--- and the entanglement is swapped to the extreme photons entering Alice's and Bob's measurement devices. Only in this successful event, Bob selects his measurement setting. In this way, unheralded signals can be withdrawn without opening the detection loophole. For the purpose of key generation, Bob's device contemplates a third input setting ($y=2$) matching one of Alice's test bases.
	\label{fig:QA}
\end{figure}

As an example, a quantum non-demolition measurement for the non-destructive detection of a photonic qubit may play the role of a heralding mechanism, and significant progress has recently been reported in this direction~\cite{niemietz2021nondestructive}. Alternatively, another solution are the so-called ``qubit amplifiers" (QAs)~\cite{Gisin_QA,Pitkanen,Curty,Zapatero,Meyer-Scott,Kolodynski}. Leaving the technicalities aside, a QA essentially is a teleportation gate located at one party's site, such that a successful teleportation locally warns that party about the arrival of a photon. An example of QA-assisted DI-QKD is illustrated in Fig.~\ref{fig:QA}.

Similarly, a related approach more symmetric with respect to both parties is to deliver entanglement via entanglement swapping~\cite{zukowski1993event}. In this case, local quantum systems at both sites are respectively entangled with photonic states sent to a central node where the swapping occurs. From there, a successful entanglement distribution is communicated back to both parties. As discussed later in Section~\ref{Implementations}, this approach provides the foundation for the recent memory-based DI-QKD experiments~\cite{Nadlinger,Zhang}, where the heralded entanglement is established between long-lived matter-based quantum memories.

Undoubtedly, heralding mechanisms seem to be a mandatory breakthrough to enlarge the distance potentially covered by DI-QKD. However, various aspects must still be improved for their applicability. In particular, a general bottleneck of heralding schemes is that, with current technology, very long DI-QKD sessions would be required to gather the data block sizes necessary to deliver a positive finite key length at relevant distances~\cite{Zapatero}. Also, in an all-photonic implementation, the performance of the heralding scheme is limited when one considers practical entanglement sources that sometimes emit vacuum pulses or multiple photon pairs, like e.g. those based on spontaneous parametric down-conversion (SPDC)~\cite{PDC1,PDC2,Caprara,Seshadreesan,Zapatero}. As for the actually implemented memory-based approach (on which we elaborate later), the use of SPDC sources is circumvented by exploiting ion-photon or atom-photon entanglement, but the time issue is further magnified by quantum memory inefficiencies.\\

\noindent \textbf{Protocol improvements and variants} \\

To relax the requirements of DI-QKD, various modifications of the original CHSH-based protocol have been proposed as well, which must in any case be combined with the use of a heralding mechanism to achieve long distances. In what follows, we briefly discuss some of them. For this purpose, it is convenient to quantify the Bell violation with the CHSH value,
\begin{equation}\label{S}
S=\langle{a_{0}b_{0}}\rangle+\langle{a_{0}b_{1}}\rangle+\langle{a_{1}b_{0}}\rangle-\langle{a_{1}b_{1}}\rangle,
\end{equation}
where the correlators $\langle{a_{x}b_{y}}\rangle$ are defined as  $\langle{a_{x}b_{y}}\rangle=p(a=b|x,y)-p(a\neq{}b|x,y)$. Straightforwardly, the CHSH winning probability and the CHSH value are related as $\omega=S/8+1/2$, such that the CHSH inequality and the Tsirelson's bound respectively read $S\leq{}2$ and $S\leq{}2\sqrt{2}$.

This said, a possible improvement is the addition of a noisy preprocessing step~\cite{Ho} after the estimation of $S$, where Alice independently flips each sifted key bit with a certain probability. Remarkably, it was shown in~\cite{Ho} that noisy preprocessing can decrease the minimum detection efficiency with an SPDC source to $83.2\%$, if a multi-mode approach is deployed for photon counting with threshold detectors, and the refined data-processing of~\cite{Ma} is incorporated for EC. This suggests that the development of perfect entanglement sources might not be a priority for all-photonic DI-QKD.

Another possibility to slightly decrease the minimum detection efficiency and enhance the secret key rate is to test asymmetric~\cite{Woodhead} or generalized~\cite{Sekatski} CHSH inequalities, where one suitably modifies $S$ by incorporating non-unit weights in the first and the second pairs of correlators appearing in Eq.~(\ref{S}). Given an accurate model of the devices' behaviour (which does not enter the security analysis and hence does not compromise the DI feature), the free weights can be optimized in a model-dependent way to enhance the bounds on the secret key rate. In particular, combining this technique with the noisy preprocessing idea, the authors of~\cite{Woodhead} report a minimum detection efficiency of $82.6\%$ for DI-QKD. This value is attained by considering partially entangled states with an optimized bias and optimized measurement settings as well. Indeed, the possibility of reducing the critical efficiency of a Bell test by using non-maximally entangled states was already pointed out in~\cite{Eberhard}. Similarly, it was shown in~\cite{Caprara} that, with unit efficiency threshold detectors, the maximum CHSH violation accessible to an SPDC source without a heralding mechanism ($S\sim{2.35}$) is attained with non-maximally entangled states.

In a similar fashion, instead of simply estimating the CHSH value, one could use the test rounds of the protocol to fully characterize the input-output statistics $p(a,b|x,y)$ of the measurement devices with the available data. It has been shown in~\cite{Tan,Brown} that this finer-grained analysis (which in fact is not restricted to the CHSH setting, but also applicable to general Bell inequalities) allows to tighten the security bounds and significantly lower the minimum detection efficiency. To be precise, the formerly established threshold of 90.9\%~\cite{Pironio,Ma} is reduced to 84\% in~\cite{Brown} without noisy preprocessing. A related approach exploiting both the complete statistics and the noisy preprocessing is given in~\cite{xu2021device}, where a DI-QKD protocol with a random post-selection technique is proposed. This technique, originally presented in~\cite{delaTorre} and limited to the collective attack model (discussed in Section~\ref{Security}), is essential for the proof-of-principle experiment reported in~\cite{Liu}.


Also, a suggested protocol modification to enhance the noise-tolerance is to randomly alternate between two different key-generating bases~\cite{Schwonnek,Masini}. Precisely, both of Alice's test bases are used for key generation in this proposal, and Bob incorporates a new measurement setting accordingly. The intuition behind this idea is that Eve cannot tune her attack to simultaneously maximize her information gain of the measurement outcomes in both (incompatible) key bases. Importantly, in the high noise regime, this additional difficulty for Eve compensates the extra sifting arising from the basis-mismatch probability in the key rounds. Notwithstanding, although the random-key-basis idea increases the tolerated quantum bit error rate (QBER) between the parties' key strings by $\sim{1}\%$ within the typical depolarising-noise model~\cite{Pironio} ---relevant, for instance, in the experiment reported in~\cite{Zhang}---, it might not be advantageous in the limited-efficiency model (which we recall is standard for benchmarking the minimum tolerated detection efficiency of the different protocol variants).

Lastly, the authors of~\cite{Masini} also consider the original single-key-basis protocol with deterministic assignments of the lost signals~\cite{Pironio}, and combine the noisy preprocessing idea with the possibility of taking into account the expected value of Alice's key generation outcome (in addition to the CHSH value) for the parameter estimation. Using this approach and deliberately contemplating partially entangled states, full optimization of the bias of the latter and the measurement settings allows to decrease the minimum detection efficiency to $80.3\%$.

Needless to say, in all these protocol variants, the minimum detection efficiency is expected to increase if one takes into account any additional form of noise not present in the limited-efficiency model (see e.g.~\cite{Woodhead}). In this respect, it has been shown in~\cite{Cabello1,Cabello2} that higher-dimensional Bell inequalities offer lower minimum efficiencies and better robustness to noise than the CHSH inequality. Note, however, that solutions like these rely on the possibility of entangling two particles in higher-dimensional degrees of freedom, which is a more complex task from an experimental point of view. But of course, technology is improving, and these ideas may eventually become a feasible experimental alternative for DI-QKD.\\

\noindent \textbf{Multipartite DI-QKD} \\

To finish this section, it is worth mentioning that the alternative of multipartite DI-QKD ---sometimes termed DI conference key agreement (CKA)~\cite{Holz}--- has also been explored in theory. The goal here is to distribute an equal secret key among $n>2$ users. In~\cite{Holz}, a non-trivial generalization of the CHSH inequality is devised for the implementation of a DI-CKA protocol with the $n$-partite Greenberger-Horne-Zeilinger (GHZ) state~\cite{GHZ} ---which is a natural extension of a Bell state to the multipartite setting. Considering a depolarising-noise model for every qubit subsystem, the authors of~\cite{Holz} show that, in the low noise regime, their DI-CKA protocol reaches larger asymptotic key rates than the combination of multiple bipartite DI-QKD protocols (as long as Alice cannot perform all the bipartite protocols at the same time).
\section{Security of DI-QKD}\label{Security}
So far, we have mentioned the term ``security'' (of a DI-QKD protocol) without explicitly saying what is meant by that. Informally, a DI-QKD protocol is secure if, for any device that implements the protocol, either a ``good key'' is produced, or the protocol aborts with high probability. A ``good key'' refers to one that is sufficiently close to a key with the following two properties: (a)~The key is unknown to the adversary (b)~Alice and Bob hold the same key. 
    These statements can be made precise by formal definitions that capture the above and ensure that the produced key can be used freely in subsequent applications; see~\cite{PortmannRenner,arnon2020device,Wolf2021} for didactic and rigorous explanations.  
    To prove the security of a protocol, one must show that the considered protocol fulfils the requirements set by the security definitions. 
    
    For most protocols, the main theoretical challenge when proving security is to provide a lower bound on a quantity called the smooth conditional min-entropy~\cite{Renner05,Tomamichel10,Tomamichel15} $H_{\min}^{\varepsilon}(\mathbf{A}|E)$, where $\mathbf{A}$ stands for Alice's sifted data, $E$ denotes the information that Eve gathers on $\mathbf{A}$ (including all knowledge leaked to her during the execution of the protocol) and $\varepsilon\in(0,1)$ is a security parameter. 
    Roughly speaking, the smooth conditional min-entropy quantifies the (lack of) knowledge that the adversary has about the sifted data.  
    Once a sufficiently high lower bound on the smooth conditional min-entropy is derived, the PA step of the protocol, in which a strong quantum-proof extractor~\cite{RennerKonig05,KonigTerhal08,FehrSchaffner08,De12} is being applied, guarantees that the final key, produced from $\mathbf{A}$, is indeed unknown to the adversary. 
    Thus, for the rest of this section, we mainly focus on explaining the main techniques used in order to lower bound  $H_{\min}^{\varepsilon}(\mathbf{A}|E)$. 
    
    For simplicity, let us first assume that the devices behave in an identical and independent manner in each round of the protocol, i.e.,  that they use the same set of measurements on the same state in each round. This is also called the IID (standing for independent and identically distributed) assumption or the collective attack model. In this case, the quantum asymptotic equipartition property (AEP)~\cite{Tomamichel09} tells us that, to the first order in the total number of rounds~$n$,
    \begin{equation}
    H_{\min}^{\varepsilon}(\mathbf{A}|E) \approx n H(A|E),
    \end{equation}
    where $H(A|E)$ is the conditional von Neumann entropy ``produced'' in a single round.  
    Informally, we can say that the total amount of entropy in the system is the sum of its parts.
    Note that the transition from the smooth conditional min-entropy to the conditional von Neumann entropy, as written above, is of great importance: this is what allows one to get bounds on the amount of randomness ---and via this, on the key rate--- which are tight to the first order in $n$. 
    Under the IID assumption, it remains to find a (tight as possible) lower bound on $H(A|E)$, as a function of the winning probability in the considered non-local game. For the CHSH game, this bound was first derived in~\cite{Pironio} considering the asymptotic regime. Alternatively, for other Bell inequalities or when utilizing the full statistics, one may use the approach of e.g.~\cite{Brown} to find a lower bound on the von Neumann entropy.
    
    Of course, making the IID assumption in the DI setting is not justified: there is no reason for the device to behave in an independent and identical way in every round of the protocol. What we do know, however, is that the entire operation of the device is sequential, meaning that the protocol is executed one round after the other. Hence, the operations of the device in one round may impact the following rounds, while future rounds cannot affect the past.
    
    This sequential structure lies at the heart of the techniques used to prove the security of DI-QKD protocols against general attacks. In particular, the entropy accumulation theorem (EAT)~\cite{Dupuis2020,Dupuis2019,Metger2022} can be seen as an extension of the above mentioned AEP, where instead of the IID assumption the sequential structure (satisfying certain conditions) is used. As the AEP, the EAT states that $H_{\min}^{\varepsilon}(\mathbf{A}|E) \approx n H(A|E)$. Now, however,  $H(A|E)$ should be understood as the ``worst-case von Neumann entropy'' in a single round. 
    Then, using the EAT, the security of DI-QKD can be proven in full generality, i.e., for the strongest possible adversary~\cite{Arnon2019,Arnon2018}.
    
    Other proof techniques that are used to lower bound $H_{\min}^{\varepsilon}(\mathbf{A}|E)$, such as those based on quantum probability estimators~\cite{Knill2018,Knill2020} or the complementarity approach~\cite{Zhang21}, also exploit the sequential structure of the protocols. 
    An exception is the approach of~\cite{Jain2020,Vidick2017}, that considers ``parallel-input'' protocols in which the  device may perform many rounds all at once, and thus the sequential structure is broken. 

Importantly, quantifying Eve's uncertainty about Alice's key for PA purposes is not the only necessary task. As Alice's and Bob's final keys should be identical, we also need to make sure that their bit strings match in the key generation rounds. For this, Alice and Bob need to employ a classical EC protocol on their data, during which some information is transferred between Alice and Bob and, via this, leaks to Eve and increases her knowledge about the data. Therefore, we wish to minimize the amount of communication needed in this step. The quantity that allows one to calculate the minimum amount of communication required for successful EC is the QBER, which we recall is defined as the probability that Alice's and Bob's measurement outcomes are different in the key generation rounds. When optimizing a DI-QKD protocol for a specific setup, one should consider the expected QBER in the actual experiment and choose the parameters of the EC protocol accordingly.

\begin{figure}[!htbp]
\centering 
	\includegraphics[width=7.4cm,height=15cm]{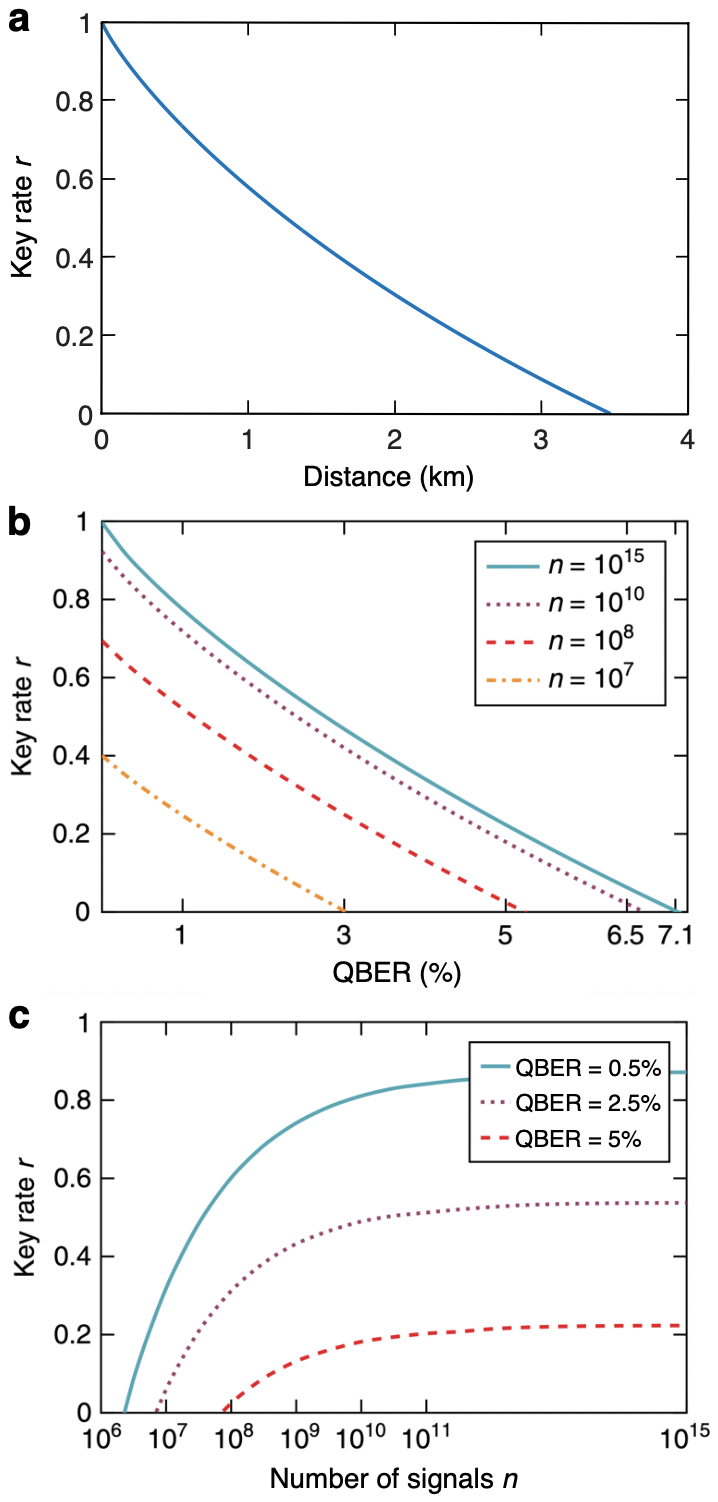} 
	\caption{{\bf Secret key rate performance of DI-QKD.} {\bf a} Asymptotic DI Devetak-Winter secret key rate as a function of the distance~\cite{Devetak2005,Kaur2020,Arnon2021}. An idealized scenario is assumed where Alice and Bob implement unit efficiency measurements on their shares of a perfect Bell pair, delivered by an entanglement source equidistant from both of them. Moreover, the standard approach where undetected signals are mapped to a fixed outcome is presumed for the closure of the detection loophole. Noticeably, even in this idealized setup, the absence of a heralding mechanism implies that the distance covered by DI-QKD is below 3.5 km, considering an attenuation coefficient of 0.2 dB/km (referred to the third telecom window). The limited-efficiency model is considered for the channel loss~\cite{Pironio}. {\bf b, c} Finite secret key rate as a function of the QBER and the number of transmitted signals, $n$, assuming a depolarising-noise model for a perfect Bell pair and no losses of any kind. For illustration purposes, some typical finite-key parameters are selected. Both figures, b and c, are taken from~\cite{Arnon2018}.}
\label{fig:rates}
\end{figure}

A few last remarks regarding key rates are in order. 
We did not give explicit equations for the secret key rates achieved in the above works, because these depend on many parameters and are not very informative by themselves. 
Asymptotically though, the works that exploit the sequential structure of the protocols achieve a key rate that matches the ``DI Devetak-Winter'' key rate~\cite{Devetak2005,Kaur2020,Arnon2021}, i.e., the one that is also achieved under the IID assumption, but without making this assumption in the first place (see Fig.~\ref{fig:rates}~a). 
Some of the cited works (such as e.g.~\cite{Arnon2018,Schwonnek,Bhavsar2021}) and various others derive explicit bounds in the finite key regime, that is to say, for any finite number of rounds. For completeness, Fig.~\ref{fig:rates}~b and Fig.~\ref{fig:rates}~c illustrate the finite key security bounds obtained in~\cite{Arnon2018}.

\section{DI-QKD implementations}\label{Implementations}
In a nutshell, the implementation of DI-QKD requires an experimental platform that distributes entanglement with high fidelity, detection efficiency, and rate, over distances that are relevant for cryptography. More specifically, a sufficiently entangled state and highly efficient appropriate measurements are required to largely violate a Bell inequality while achieving a low QBER, which is mandatory for key generation. With current technology, it remains a major challenge to simultaneously achieve all these conditions, however, the first proof-of-principle demonstrations have recently been reported~\cite{Nadlinger, Zhang, Liu}.

Photons are the physical system of choice to distribute entanglement. For instance, one can encode a quantum state in the polarization or time degrees of freedom of a photon, and guide it to a distant location using an optical fibre~\cite{gisin2007quantum}.

For the purpose of DI-QKD, we can distinguish two main implementation categories: all-photonic setups in which the qubits are encoded using photons only, or memory-based setups employing long-lived matter states to generate and store heralded entanglement. Both types of setups have different advantages and face different implementation challenges.

All-photonic setups can generate and distribute entangled states at high rates and with high fidelities, which makes this approach promising for cryptographic applications. For instance, in the ubiquitous example of SPDC sources, the generation of entanglement originates from a single-photon conversion process~\cite{PDC1}, which is very convenient for a high rate. In fact, minimizing photon-coupling losses and being capable of fastly and randomly switching between different measurement bases enabled the violation of Bell inequalities while closing the detection loophole~\cite{giustina2013bell,christensen2013detection}, even over distances sufficient to simultaneously close the locality loophole~\cite{Giustina, Shalm}. Despite the preliminary success of photonic Bell tests though, single-photon detectors and distribution through fibres limit the global detection efficiency and hence the distance of all-photonic DI-QKD, unless the heralded approach presented in Section~\ref{Protocols} is deployed. Nevertheless, DI-QKD based on all-photonic heralding schemes has not been realized so far, although all the necessary experimental methods have been demonstrated individually.

On the other hand, memory-based approaches generate heralded entanglement between two matter-based quantum memories. Such schemes have been demonstrated using different systems as quantum memories and allow to close the detection loophole~\cite{Hensen,Rosenfeld}. Essential here is, however, to realize efficient light-matter interfaces and high-speed entanglement generation procedures to shorten the necessary duration of a potential DI-QKD session.\\

\noindent \textbf{Photonic-based implementations}\\

The most efficient photonic experiment reported until recently~\cite{liu2021device} (deployed for DI randomness expansion) reached a single-photon detection efficiency of around 84\%. At the time the experiment was carried out, the CHSH-based protocol variants that lower the detection efficiency below that value had not been proposed yet (e.g.~\cite{Ho,Woodhead,Brown}), and hence it was considered that the system did not keep up with the requirements of DI-QKD in practice.

\begin{figure*}[thbp]
	\centering
	\resizebox{15cm}{!}{\includegraphics{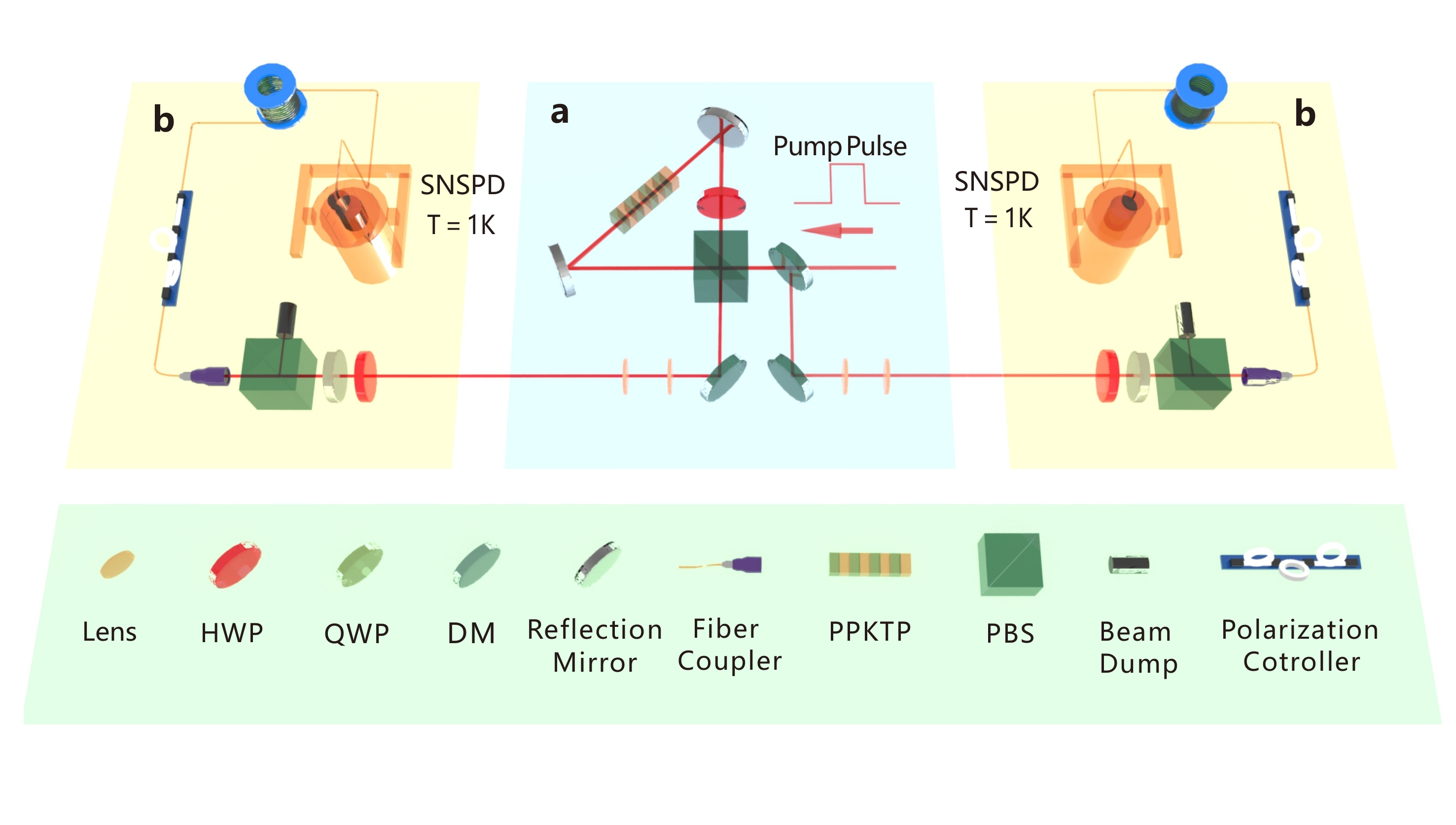}}\vspace{-0.5cm}
	\caption{{\bf Schematic of the all-photonic DI-QKD implementation reported in~\cite{Liu}.} {\bf a} Entanglement source. For the creation of pairs of entangled photons, light pulses of 10~ns are injected at a repetition pulse rate of 2~MHz into a periodically poled potassium titanyl phosphate (PPKTP) crystal in a Sagnac loop to generate polarization-entangled photon pairs. The two photons of an entangled pair at 1560~nm travel in opposite directions towards Alice and Bob, where they are subject to polarization projection measurements. {\bf b} Alice's and Bob's QKD receivers. In her (his) measurement site, Alice (Bob) uses a set of HWP and QWP to project the polarization of the incoming single photons into a pre-determined measurement basis. After being collected into the fibre, the photons travel through a certain length of fibre and then are detected by a superconducting nanowire single-photon detector (SNSPD) operating at 1K. HWP -- half-wave plate; QWP -- quarter-wave plate; DM -- dichroic mirror; PBS -- polarizing beam splitter. The figure is taken from~\cite{Liu}.}
	\label{fig:setup}
\end{figure*}

Very recently, the combination of various experimental simplifications with the post-selection technique of~\cite{xu2021device} enabled the realization of a proof-of-principle all-photonic DI-QKD experiment without a heralding mechanism~\cite{Liu}. Specifically, the applied random post-selection technique significantly reduces the error events, leading to tolerable detection efficiencies as low as 68.5\% in the limited-efficiency model. However, this technique ---which is central to enable the delivery of a positive asymptotic key rate in the experiment--- relies on the collective attack model in a fundamental way (see~\cite{delaTorre}). Therefore, more research is needed to investigate whether it can be extended to the fully DI setting against general adversaries.

As for the experimental simplifications, the measurement settings were not randomly changed from round to round in the experiment, but rather waveplates were set manually to determine all the correlators for evaluating $S$. In addition, aiming to simulate longer distances, optical fibres with a total length of 220 m were placed behind the state analysis to avoid the otherwise required stabilisation of the polarization.

In any case, within this preliminary simplified scenario, after $2.4\times10^8$ rounds of experiment for each of the six combinations of measurement settings, $2.33\times10^{-4}$ secret key bits per pulse are expected to be achieved asymptotically. A schematic of the setup is depicted in Fig.~\ref{fig:setup}, which consists of three modules. Pairs of polarization-entangled photons at the wavelength of 1560~nm are generated probabilistically via the SPDC process in the central module (a). These pairs of photons are sent over a short distance to two receiver modules (b). In each module, approximately 100 m of fibre precede the single photon detectors where the measurements are performed to generate the raw data. The overall single-photon detection efficiencies are respectively determined to be $87.16\pm0.22\%$ and $87.82\pm0.21\%$ for Alice and Bob, which significantly surpass the record values in previous experiments with photons.

In short, despite the great progress that the experiment represents~\cite{Liu}, an all-photonic implementation delivering a positive finite key length may still require further technological improvements. On top of it, needless to say, the deployment of all-photonic heralding schemes is also a must aiming to cover relevant distances.\\

\noindent \textbf{Memory-based implementations} \\

\noindent Quantum memories with light-matter interfaces enable the generation of heralded entanglement between distant locations. For this, the memories first emit a photon to generate entanglement between light and matter~\cite{blinov2004observation, matsukevich2004quantum, volz2006observation} or interact with incoming light in a state dependent manner~\cite{julsgaard2004experimental,wilk2007single}. Two distant memories can then be entangled by using, for example, heralded storage of an entangled photon pair, entanglement swapping from two light-matter pairs, or enhanced light-matter interaction by resonators.
Various quantum systems are under active research to facilitate entanglement distribution, and heralded entanglement has been generated for platforms including ions~\cite{matsukevich2008bell}, atoms~\cite{hofmann2012heralded,van2021entangling}, and NV-centers~\cite{bernien2013heralded}, even over long distances and with readout speeds capable of simultaneously closing the locality loophole~\cite{Hensen, Rosenfeld}.

Heralded entanglement generation between quantum memories allows to close the detection loophole in a Bell test regardless of channel loss, however, DI-QKD is more demanding than this alone~\cite{farkas2021bell}. The current challenge is to distribute high-quality entangled states and, at the same time, to achieve entanglement rates over distances that are relevant for cryptography. The highest entanglement fidelity reported so far equals 96\%, employing two distant ion based quantum memories~\cite{stephenson2020high, Nadlinger}. These fidelities are not fundamentally limited though, and could be increased by further reducing state generation or readout errors. Generation rates of high quality entangled states up to 100 Hz have been reported~\cite{humphreys2018deterministic, stephenson2020high,stockill2017phase}, and higher rates could still be achieved employing cavities to improve photon collection efficiencies~\cite{schupp2021interface, brekenfeld2020quantum}. Moreover, single-photon interference provides a promising venue for entanglement generation protocols~\cite{cabrillo1999creation, pompili2021realization, yu2020entanglement}, although so far only at the price of reducing the quality of the entanglement.

While the memory-based heralding schemes do not have a fundamental distance limitation, due to absorption the entanglement generation rate decreases exponentially over the distance. To minimize channel loss and hence maximize the rate using optical fibres, operation at telecom wavelengths is indispensable. Indeed, entanglement between quantum memories has already been distributed over tens of km employing quantum frequency conversion~\cite{yu2020entanglement,van2021entangling} or absorptive quantum memories~\cite{lago2021telecom,liu2021heralded}.

\begin{figure*}
\includegraphics[scale=1]{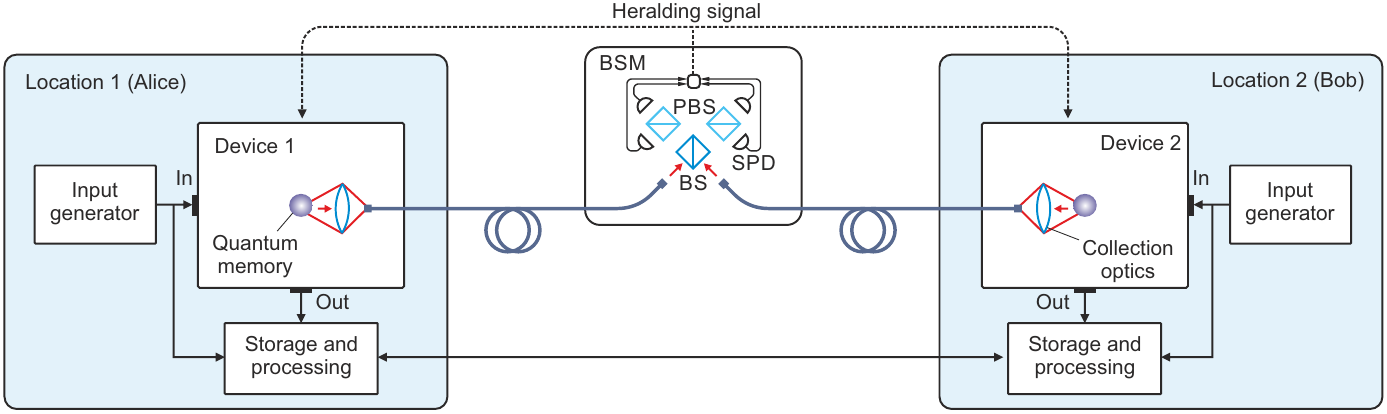}
\caption{\textbf{Schematic of the memory-based DI-QKD implementations reported in~\cite{Nadlinger,Zhang}.} Alice and Bob are situated at distant locations and equipped with an input generator, storage and process unit, and a device containing a single ion or atom quantum memory. The memories are entangled as follows. First, via spontaneous decay entangled atom-photon pairs are generated in each device. The photons are collected into fibres that guide them to a Bell-state measurement (BSM) setup. There, they interfere at a beam splitter (BS) where in each output port the photon polarization is analysed with a polarizing BS (PBS) and two single photon detectors (SPD). The entanglement generation tries are repeated till a photonic coincidence detection occurs and heralds shared entanglement between the quantum memories. Next, the DI-QKD protocol starts with a random seed from the input generators to select the measurement orientation, perform the measurement, and output the result. These measurement in- and outputs of every round are stored in a local storage. The storage and processing unit are connected via an authenticated classical channel to facilitate the post-processing steps of the key generation procedure.}
\label{fig:exp-memory}
\end{figure*}

\begin{figure}
\includegraphics[width=0.4\textwidth]{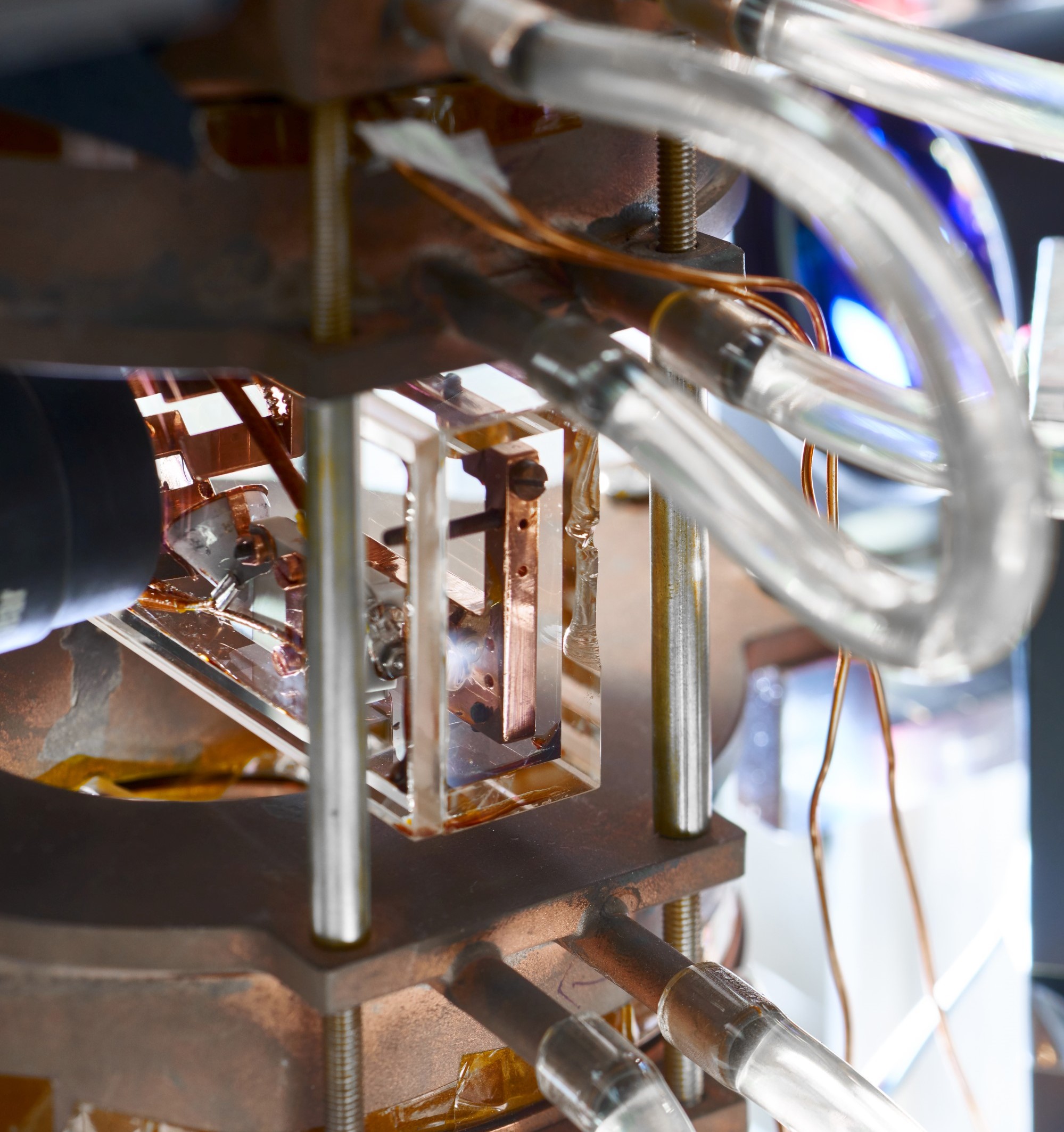}
\caption{\textbf{Picture of a single-atom trap employed in~\cite{Zhang}.} Shown is the ultra-high vacuum glass cell in which a single rubidium atom is stored. \copyright Jan Gruene (LMU).}
\label{fig:exp-memory-picture}
\end{figure}

Recently, two similar proof-of-principle implementations of memory-based DI-QKD have been reported: one using single $^{88}$Sr$^+$ ions in Oxford~\cite{Nadlinger} and the other using single $^{87}$Rb atoms in Munich~\cite{Zhang}. The experiments employ charged or neutral single atoms as quantum memories which are isolated from the environment inside ultra-high vacuum setups and spatially confined using electrically or optically induced trapping potentials, respectively. Other than the trapping techniques, the concepts used to distribute entanglement are very similar for the two implementations, and Fig.~\ref{fig:exp-memory} shows a high-level schematic of the quantum network link that is representative for both setups. Also, Fig.~\ref{fig:exp-memory-picture} shows a picture of the single-atom trap employed in \cite{Zhang}.

Each link consists of two distant quantum memories that are entangled in an event-ready scheme. First, each of the atomic spin states is entangled in a spontaneous decay with the polarization of the emitted photon. The two photons are guided with single-mode fibres to a BSM device where a joint measurement on the photons heralds the entanglement between the ions or atoms, respectively. The quantum state of the memories is measured after every heralding signal with unit detection efficiency, thereby closing the detection loophole for the Bell test.

Performance parameters of the two quantum links are listed in Table~\ref{tab:quantum-links}. The rate at which the links generate entanglement critically depends on the success probability of the entanglement generation tries and their repetition rate. The former is mainly limited by the efficiency to collect photons emitted by the memories, which is typically up to a few percent using free-space optics. Including channel and detection losses, this leads to heralding probabilities on the order of $10^{-4}$ to $10^{-6}$. For the implemented event-ready schemes, the entanglement generation tries can only be repeated after a period that allows for two-way communication between the devices and the BSM setup, thus introducing a trade-off between distance and rate. The achieved fidelities of the states shared between the distant quantum memories belong to the highest reported so far (see Table~\ref{tab:quantum-links}) and allow one to achieve positive key rates in DI-QKD protocols.

\begin{table}[]
\caption{\textbf{Performance of the quantum links.} Key parameters that characterize the quantum link performance employed in the proof-of-principle memory-based DI-QKD experiments. The line-of-sight distance between the two quantum memories is 2 and 400 m for the Oxford and Munich experiments, respectively.}
\bgroup
\def\arraystretch{1.5}%
\setlength{\tabcolsep}{10pt}
\begin{center}
\begin{tabular}{lcc} \hline
                & Oxford~\cite{Nadlinger} & Munich~\cite{Zhang}    \\ \hline
Fidelity        & 0.960(1) & 0.892(23) \\
Rate (s$^{-1}$) & 100      & 1/82      \\
Fibre length (m)    & 3.5        & 700       \\ \hline
\end{tabular}
\end{center}
\egroup
\label{tab:quantum-links}
\end{table}


As discussed in Section~\ref{Assumptions}, proper isolation of the users' locations is a way to practically close the locality loophole in the Bell test. The generation of entanglement, however, requires the photonic channel, i.e., a connection between the quantum memories, and hence opens a back door to the outside environment. Therefore, after distributing the entanglement, this link should be disconnected. For this, in the Oxford experiment, the ions are moved out of the focus of the optics and hence prevent fluorescence from coupling into the fibre leaving the laboratory. In the Munich experiment, a shutter closes the atomic fluorescence light path leaving the laboratory. Moreover, before repositioning the ions or opening the shutter, the atomic states are scrambled or the atoms are ejected from the trap to avoid information leakage to the environment when reconnecting the quantum link. These processes must be well-characterized, to ideally quantify any possible information leakage and account for it in the security proof.

The key distribution capability of the ion-based system was evaluated by generating a secret key between Alice and Bob. For this, a finite-key security analysis was considered, together with a protocol that implemented EC, authentication and PA. A total of $1.5\times10^6$ Bell pairs were generated during a period of 7.9~h, achieving a CHSH value of $S=2.677(6)$ and a QBER of $Q=0.0144(2)$. After a post-processing time of 5~min, the protocol generated $95\,884$ shared secret bits (i.e., 0.064 bits per entanglement generation event), while only 256 bits were consumed during the key generation process.

For the atom-based system, an asymptotic security analysis was made, since the generation of a key secure under finite statistics would have taken months of quantum communication. In particular, over a period of 75~h, a total of $3,342$ entanglement generation rounds were executed, observing a CHSH value of $S=2.578(75)$ and a QBER of $Q=0.078(9)$. This translates into an asymptotic secret key rate of 0.07 bits per entanglement generation event (compared to the maximum of 0.25 for the protocol used).

Single photons with respective wavelengths of 422 and 780~nm distributed the entanglement, corresponding to an attenuation loss of a factor two for every 100 and 700~m of optical fibre in each case. To achieve entanglement distribution over more than a few kilometres of fibre, it is required to operate at low-loss telecom wavelengths. In this regard, conversion of light from 422~nm to the telecom regime was recently achieved in a first demonstration setup~\cite{wright2018two}. Moreover, fully polarization-preserving quantum frequency conversion was implemented in the atom-based experiment, which converts the wavelength from 780~nm to telecom wavelength for a single photon while maintaining its quantum state~\cite{ikuta2018polarization}. The high conversion efficiency of about 57\% allowed to distribute entanglement of atomic quantum memories over tens of kilometres of fibre~\cite{van2021entangling}. However, besides providing an even lower entanglement generation rate, the achieved fidelity would not be sufficient for DI-QKD yet, due to the decoherence of the atomic quantum memories.

\section{Outlook}\label{Outlook}

    
    Over the last years, crucial theoretical and experimental advancements have been made, which enabled the first proof-of-principle QKD demonstrations in the DI setting. For practical implementations, however, more effort is required. 
    
    On the theory side, the analysis of the studied schemes is tight when considering CHSH-based protocols. To improve the performance, more sophisticated protocols including e.g. two-way classical communication in the post-processing stage~\cite{advantage} or exploiting higher-dimensional Bell inequalities~\cite{Cabello1,Cabello2} might be an option, though they seem to be challenging at the moment.
    
    On the experimental side, all-photonic approaches including heralding schemes are a promising venue towards DI-QKD over tens of kilometres, where fully integrated photonics could further improve the efficiencies of the devices. In memory-based implementations, the achieved entanglement fidelities between distant matter qubits allow for positive key rates in the DI setting, but a severe challenge is to achieve high entanglement rates over distances which are relevant for key distribution. For this, using e.g. cavities could increase the efficiencies of the light-matter interfaces. Especially for event-ready schemes, parallelization of the entanglement generation tries can potentially increase the entanglement generation rate by orders of magnitude. For this, novel hardware architectures that were initially proposed for quantum simulation and computation applications could be exploited, for example, using strings of trapped ions or atom arrays~\cite{ramette2022any,dhordjevic2021entanglement}.
    
    
    Another interesting research direction is ``other forms'' of DI protocols. That is, protocols in which we do not characterize the devices but the assumptions we do make are not comparable to those described in Section~\ref{Assumptions};~\cite{MDI,TF,Metger,Thomas} being examples of such scenarios. 
    It is for the theoreticians and experimentalists together to investigate which models are of relevance and develop both the security proofs and the necessary equipment for the implementation of the upcoming protocols.
    
    On the long run, quantum networks might be employed to efficiently transfer quantum states over long distances and provide connections between quantum computers. These networks will supply shared entanglement between the nodes, which will hopefully be accessed with very high efficiency. Once all the necessary tools be available, DI-QKD will become a regular application for secure communications of the highest level.
\section{Data availability}
No datasets were generated or analysed during the current study.
    
\section{Acknowledgements}
VZ and MC acknowledge support from the Galician Regional Government (consolidation of Research Units: AtlantTIC), the Spanish Ministry of Economy and Competitiveness (MINECO), the Fondo Europeo de Desarrollo Regional (FEDER) through Grant No. PID2020-118178RB-C21, Cisco Systems Inc., and MICINN ---with funding from the European Union NextGenerationEU (PRTR-C17.I1)--- and the Galician Regional Government ---with own funding--- through the “Planes Complementarios de I+D+I con las Comunidades Autonomas” in Quantum Communication. RAF was generously supported by the Peter and Patricia Gruber Award, the Daniel E. Koshland Career Development Chair and by the Israel Science Foundation (ISF), and the Directorate for Defense Research and Development (DDR\&D), grant No. 3426/21. TvL and HW acknowledge funding by the German Federal Ministry of Education and Research (Bundesministerium f{\"u}r Bildung und Forschung (BMBF)) within the project QR.X (Contract No. 16KISQ002) and the Deutsche Forschungsgemeinschaft (DFG, German Research Foundation) under Germany’s Excellence Strategy – EXC-2111 – 390814868. WZL and QZ acknowledge support from the National Natural Science Foundation of China (No. T2125010) and the Shanghai Municipal Science and Technology Major Project (No. 2019SHZDZX01).\\

\section{Author contributions}
All authors discussed the work together and decided the structure and contents of the paper. VZ, TvL, RAF and WZL wrote the manuscript, with feedback from the rest. All authors critically read it and revised it. MC supervised the project.

\section{Competing interests}
The authors declare no competing interests.
\section{References}

\end{document}